\documentclass[journal,onecolumn]{IEEEtran}

\usepackage{amsmath,amssymb,graphicx,wrapfig}

\usepackage{cite}
\begin{document}

\title{Assessing Certainty of  Activation or Inactivation in  Test-Retest fMRI Studies}   

\author{Ranjan Maitra\thanks{R. Maitra is with the Department of Statistics, Iowa State University, Ames IA  50011, USA.}}



\maketitle
\begin{abstract}
Functional Magnetic Resonance Imaging~(fMRI) is widely used to study 
activation in the human brain. In most cases, data are commonly used 
to construct activation maps corresponding to a given
paradigm. Results can be very variable, hence
quantifying certainty of identified activation and inactivation over
studies is important. This paper provides a model-based approach to 
certainty estimation from data acquired over several replicates of the same
experimental paradigm. Specifically, the $p$-values derived from the
statistical analysis of the data are explicitly modeled as a mixture
of their underlying distributions; thus, unlike 
methodology currently in use, there is no subjective thresholding required
in the estimation process. The parameters 
governing the mixture model are  easily obtained by the principle of
maximum likelihood. Further, the estimates can also be used to
optimally identify voxel-specific  activation regions along with their
corresponding certainty measures. The methodology is applied to a
study involving a motor paradigm performed on a  single subject several
times over a period of two months. Simulation experiments
used to calibrate performance of the method are promising. The
methodology is also seen to be robust in determining areas of
activation and their corresponding certainties.

\end{abstract}

\maketitle



\section{Introduction}
\label{introduction}
Functional Magnetic Resonance Imaging~(fMRI) has become an extremely
popular noninvasive imaging modality for understanding human cognitive
and motor functions. 
The main goal of fMRI is to identify regions of the brain that are
activated by a given stimulus or while performing some task, but high
variability among replicated studies often leads to inconsistent
results, causing concern among researchers (see, for instance,
in Buchsbaum et al., (2005), 
Derrfuss et al., 
(2005), Ridderinkhof et al (2004), or Uttal (2001). There are a number of
factors that affect  
the identification of activated voxels. A typical fMRI paradigm
consists of the application of a stimulus or performance of a
cognitive or motor task over time.  Any neural stimulus passes through
the so-called hemodynamic filter~(Maitra et al., 2002), resulting in a
several-seconds delay before the blood-oxygen-level-dependent~(BOLD)
response occurs. Other factors also affect the acquired
data~(Genovese et al., 1997). For example, the cardiac and respiratory
motion of a subject may result 
in physiological variation, giving rise to flow-artifacts which may need
to be monitored or digitally filtered~(Biswal et al., 1996). Subjects also often exhibit
voluntary, involuntary and/or stimulus-correlated motion during
scans~(Hajnal et al., 1994).  Another factor
is scanner variability which is essentially controlled through effective 
quality control programs. Most signal differences between  
activated and control or resting states are small, typically on
the order of 1--5\%~(Chen and Small, 2007), and sub-pixel motions can induce
large apparent signal changes and result in the detection of false
positives. Therefore, fMRI data are subjected to image registration
algorithms which align the sequence of images to sub-pixel
accuracy~(Wood et al., 1998). The pre-processing of data improves
the quality of acquired fMRI data, but identified regions of
activation still vary from one replication to the other. This
variability needs to be quantified in order to determine regions of
activation with precision and accuracy~(McGonigle et al., 2000; Noll
et al., 1997; Wei et al., 2004).  

Repeatability of results across multiple studies is one way of
assessing variability and measures that calibrate repeatability are
called \textit{reliability measures}. Many authors working in the area of
fMRI image variability used the term \textit{reliability} to describe
the extent to which activation was consistently identified in multiple
fMRI images. However, there is another, perhaps more useful, quantity
of interest to practitioners: quantitation of the true status of
voxels identified as activated or inactivated. 
Measures that attempt to quantify the probability of the
true status of a voxel given its identified state are more correctly
termed measures of confidence or \textit{certainty} even though these
were also introduced, perhaps confusingly, as reliability measures by
earlier authors that included me. In  this paper, I will move towards
adopting the nomenclature of certainty in these contexts in order to
better distinguish it from simple reliability. 
But before proceeding further, I specify that I use the term
``replication'' to denote the repetition of the task or experimental
condition to study variability. These replications are necessarily 
independent and, in the context of single-subject studies, occur
on different scanning sessions, reasonably separated in time. 

The issue of quantifying variability (whether reliability or
certainty) of activation has interested
researchers in two different frameworks. The first case involves the
analysis of grouped fMRI data, which arise when 
multiple subjects are studied under multiple stimulus or task-performance
levels, {\em eg.,} fMRI data acquired while subjecting multiple volunteers
to noxious painful stimuli at several graded levels. 
I will refer to these stimulus or task levels as experimental conditions.
The second scenario, which is the focus of this paper, is the
test-retest case, where replicated fMRI data are acquired on the same
subject under the same experimental condition. 

For grouped fMRI data, the goal is to determine where the effect of the 
stimulus is larger than subject-to-subject variation.
Reliability of activation in response to stimulus has been
quantified in terms of the 
intra-class correlation~(ICC), which is calculated using voxels
identified as activated in each subject after thresholding separately
for each combination of experimental condition and subject~(Aron et
al. (2006); Fern\'{a}ndez et al., (2003); Friedman et al., (2008); Manoach
et al. (2001); Miezin et al. (2000); Raemekers et al. (2007), Sprecht et
al. (2003)). The ICC~(Shrout and Fleiss, 1979; Koch, 1982; McGraw and
Wong, 1996) provides a measure of correlation or
conformity between regions identified as activated in multiple
subjects under two or more experimental replications and/or
conditions. Thus it is inapplicable to the test-retest framework on a
single subject considered in this paper.   

For test-retest, Rombouts~et  al.~(1998) and Machielsen et
al.~(2000) have proposed a global reliability measure of the percent overlap in
voxels identified as activated between any two experimental
replications. For any two replications (say, $j$ and $m$), this
measure is calculated as 
$R_{jm} = 2V_{jm}/(V_j + V_m)$, where $V_{jm}$ is the number of
three-dimensional image voxels identified as activated in both the
$j$th and $m$th replications, and
$V_j$ and $V_m$ represent the 
number of voxels identified as activated, separately in the
$j$th and $m$th replicated experiments, respectively. $R_{jm}$ takes a
value between 0 and 
1, representing no to perfect overlap in identified activation at the
two ends of the scale. 

The percent overlap measure $R_{jm}$ provides a measurement
of the agreement in activation between any two replications,
but it is sensitive to the method of identifying activation, unusable for voxel-level analysis, and awkward for more than two replicates.
To illustrate $R_{jm}$ sensitivity to method, consider a procedure that liberally identifies activation~(\textit{eg.}, 
a naive testing approach with no correction for multiple
testing), the denominator
$V_j + V_m$ would be large 
so that small disagreement in the voxels identified
as activated would have very little impact on $R_{jm}$. 
In contrast, small differences in $V_{jm}$ would severely affect $R_{jm}$
when $V_j + V_m$ is small, as expected under a conservative method 
(\textit{eg.}, testing with the Bonferroni correction for multiple testing).
Another shortcoming is that $R_{jm}$ is a global 
measure of agreement between 
replicated experiments giving no sense of voxel-level reliability of activation.
One could compute separate $R_{jm}$ for specific brain regions, but it will never be a high-resolution measure of activation reliability. 
A third concern is that $R_{jm}$ is a reliability measure based only on the pair $(j,m)$ of experimental replicates.
When there are $M$ replicates or $M$ studies combined in a composite meta-analysis, there are $M\choose2$ overlap measures $R_{jm}$ and there
is no obvious way to combine them in a single measure of activation reliability.
Thus, there is a need for
a measure to quantify reliability or certainty of
true activation at the voxel level across an arbitrary number of replicates.
Ideally, such an assessment would be 
independent of the experimental condition and method used to identify activation. 

Some more formal statistical approaches to assessing reliability in 
the test-retest fMRI framework have been proposed as well. 
Genovese et al.~(1997) and Noll et al.~(1997) 
specified probabilities 
that voxels were correctly or incorrectly
identified as activated 
at particular thresholds of the test statistic to determine
significance of activation for a given experimental 
paradigm. Their approach modeled the total frequency~(out of $M$
replications) of a voxel identified as activated at given thresholds
in terms of a mixture of binomial distributions. 
To combine data, they
assumed independence over the thresholdings. All 
parameters, such as the mixing proportion of truly active
voxels~(denoted as $\lambda$ in their work) or probability of voxels
being correctly~($\pi_A$) or incorrectly~($\pi_I$) identified as active
were assumed to be spatially independent and estimated using maximum
likelihood~(ML) methods.  
Maitra et al.~(2002) extended their proposals by
incorporating a more accurate model of mixtures of conditional
binomial distributions, and by also generalizing $\lambda$ to be 
voxel-specific. Specifically, they let $\lambda_i$ be the probability
that the $i$th voxel is truly active. Letting $L$ be the number of
activation threshold levels (assumed without loss of generality to
be in increasing order), define $\boldsymbol{X}_i = (x_{i,1},
x_{i,2},\ldots,x_{i,L})$, where $x_{i,l}$ is the number of replications
for which the $i$th voxel is identified as activated at the $l$th
threshold. Let $\eta_{Al,l-1}$ (and $\eta_{Il,l-1}$) be the
(global) probability of a truly active  (correspondingly, truly
inactive) voxel being 
identified as activated at the $l$th threshold, given that
it has been so identified at the $(l-1)$th threshold level. Also, let
$\pi_{Al}$ (or $\pi_{Il}$) be the probability that a
truly active (correspondingly inactive) voxel is identified as
activated at the $l$th threshold. Then the likelihood function for the
$i$th voxel is  provided by
\begin{equation}
\label{maitrallhd}
\lambda_i\prod_{l=1}^L {x_{i,l-1}\choose x_{i,l}} 
\eta_{Al,l-1}^{x_{i,l}}(1-\eta_{Al,l-1})^{x_{i,l-1}-x_{i,l}} +
(1-\lambda_i)\prod_{l=1}^L {x_{i,l-1}\choose x_{i,l}}
\eta_{Il,l-1}^{x_{i,l}}(1-\eta_{Il,l-1})^{x_{i,l-1}-x_{i,l}} 
\end{equation}
where $x_{i,0}\equiv L$, $\eta_{A1,0} = \pi_{A1}$ and
$\eta_{A1,0}=\pi_{I1,0}$. 
A further generalization incorporated spatial context by
regularizing $\lambda$ through a Markov Random Field 
component in the penalty term of the estimation process. 
Estimates were obtained by maximizing the penalized
likelihood. 
Maitra et al.~(2002) 
introduced a novel approach to quantifying
certainty about the true status of voxels identified as
activated/inactivated by defining a
measure of  {\em   reliability} -- the probability of a voxel
identified as activated being truly active -- and  {\em
  anti-reliability} -- the probability of a voxel incorrectly 
identified as inactivated being active. In naming these certainty measures (as
mentioned earlier) they aligned them with the layman's
notion of reliability: trustworthiness of identified activation. 
Maitra et al.~(2002) also extended Genovese~et al.~(1997)'s approach
to provide a voxel-specific method for 
choosing the optimal threshold for detecting activation by maximizing
the ``reliability efficient frontier'' {\em i.e.}, the probability of
making a correct decision on the state of a voxel (whether activated or
inactivated) at a given threshold. 
Their emphasis was on
assessing certainty of activation and inactivation in a test-retest 
setting, 
but the method was also subsequently extended to grouped functional MR
imaging 
data
by Gullapalli~et al.~(2005).

The methodology of~Genovese et al.~(1997) and Maitra et al.~(2002) is   
implemented by obtaining a test statistic and 
thresholding it (or more commonly, its $p$-value) at different
levels. This is integral to obtaining the $x_{i,l}$s used
in~(\ref{maitrallhd}).  However, there is no clear guideline to
choosing the thresholds which is left to the 
researcher. The choice of the number $L$ and value of these thresholds is
subjective and can 
greatly impact the reliability and certainty estimates. Too few 
threshold %
levels can result in severely biased estimates, while
too many may be computationally burdensome besides having high
variability in the 
estimates. An additional issue is the subjective choice
of spacing between the thresholds, to which there is also no
satisfactory answer. In this 
paper, we reformulate the problem in order to eliminate this
requirement of 
threshold choice altogether. Specifically, we model the 
distribution of the voxel-wise $p$-value of the test statistic in
terms of a mixture of two distributions. The first component of the
mixture is the standard uniform density corresponding to the
distribution of the $p$-value under the null hypothesis of no
activation. The second is the distribution of the $p$-value when 
there is activation. While a mixture of beta distributions is
sometimes used to approximate this latter distribution 
(Pounds and Morris, 2003; Allison et al., 2002), we note that it is
possible to derive exact distributions in many standard scenarios, such as
$t$-tests. Also, the mixing proportion of the mixture component representing
the distribution of the $p$-value under activation  is the same  
as the $\lambda$ in Genovese et al.~(1997) or Maitra et
al.~(2002). Estimation is done using ML. Once 
again, optimal cut-offs can be estimated by maximizing the
reliability efficient frontier. To better reflect the fact that we are
quantifying certainty in the true status of voxels identified as
activated and inactivated, we  rename the erstwhile reliability
measure as the {\em true activation certainty} and the 
awkwardly-termed anti-reliability measure in terms of its complement
from unity, calling the latter the {\em true inactivation certainty}.
Estimates for both measures are also provided. The methodology is
applied to an experiment involving a motor paradigm that was
replicated on the same subject twelve times over the course of two
months. The performance of the suggested method is also validated via
simulation experiments over a range of replication sizes. Further, we
randomly subdivide the dataset into two subsets of six replications,
and study the robustness of the identified activation and the
corresponding true activation and inactivation certainties. We
conclude with some discussion.

\section{Theory}
\label{theory}
The $p$-value of a test statistic $T_o$ is the probability, under the
null distribution, of obtaining a more extreme value (in the direction
of the alternative) than $T_o$. For a one-sided $t$-test for the
null hypothesis $H_0: \beta = 0$ against the alternative $H_a: \beta >
0$ with $\beta$ as the regression coefficient of a general
linear model fit to the time series at a voxel, this is given by
$\mbox{I\!Pr}(t_\nu \geq T_o)$, where $\mbox{I\!Pr}$ 
abbreviates probability and $t_\nu$ denotes a $t$-distributed random
variable with $\nu$ degrees of freedom and cumulative distribution and
density functions $\Psi_\nu(\cdot)$ and $\psi_\nu(\cdot)$
respectively.  

\begin{figure*}
\includegraphics[height=3in, width=7.3in]{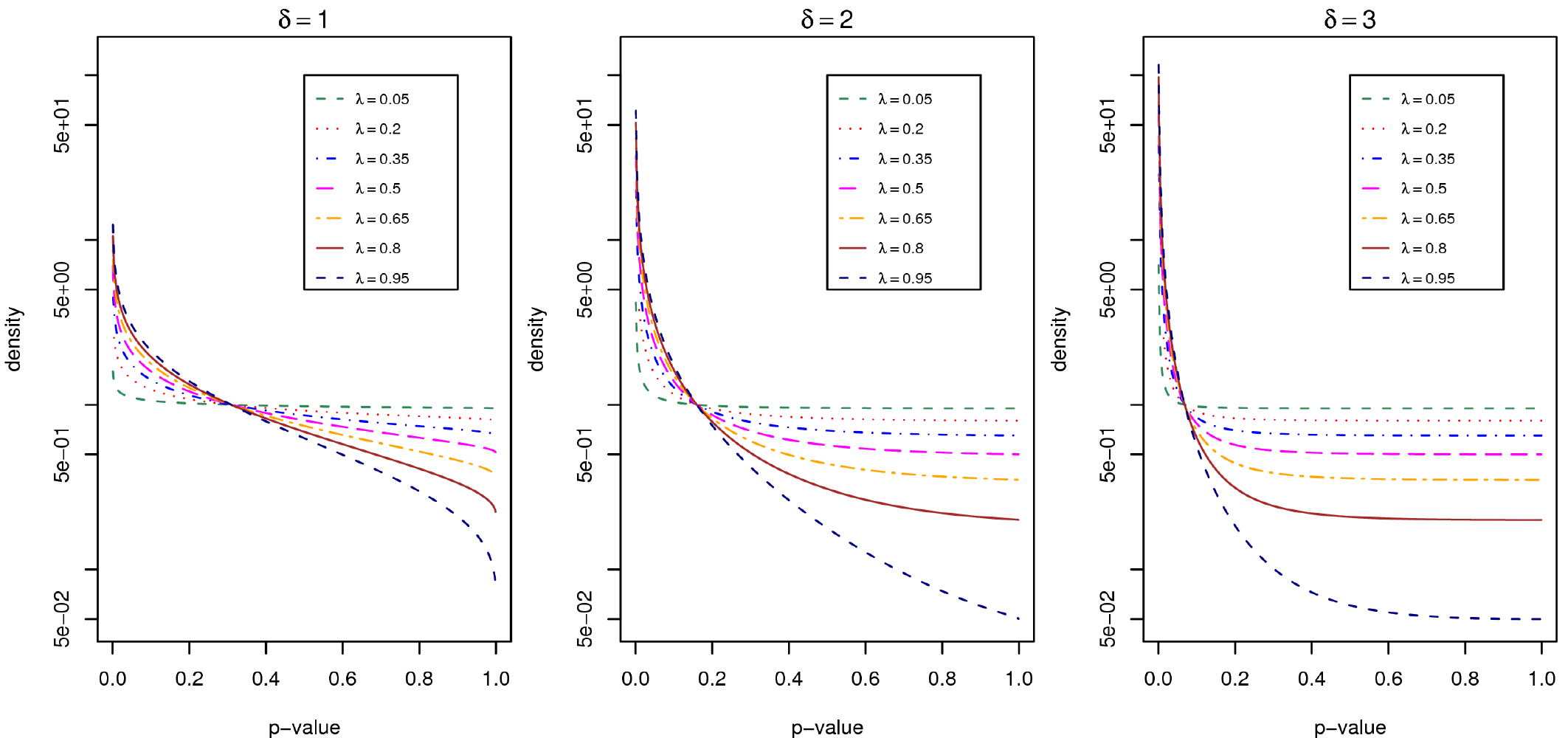} 
\caption{
Plot of the mixture density of $p$-values for different values of
$\lambda$ and $\delta$, and for $\nu = 122$.}
\label{1mixdensity}
\end{figure*}

Let $P_i$ be the $p$-value at the $i$th voxel of the $t$-statistic
with $\nu$ degrees  of freedom. Under the (null) hypothesis of no true
activation at a voxel, the $p$-value follows the standard uniform 
distribution. To see this, 
\begin{equation}
\begin{split}
\mbox{I\!Pr}(P_i\leq p \mid H_0) = \mbox{I\!Pr}\left [
  \mbox{I\!Pr}(t_\nu \geq T_o) \leq p\right ]   = \mbox{I\!Pr}\left [ \Psi_\nu(T_o) \geq 1 - p\right ]  = 1 - \mbox{I\!Pr}\left [T_o \leq \Psi^{-1}_\nu(1-p)\right ] = 1- \Psi_\nu\left[\Psi^{-1}_\nu(1-p)\right ] = p.
\end{split}
\label{pvalH0}
\end{equation}
On the other hand, under the alternative one-sided hypothesis
that the voxel is truly activated, we get 
\begin{equation}
\begin{split}
\mbox{I\!Pr}(P_i\leq p\mid H_a)  = \mbox{I\!Pr}\left [
  \mbox{I\!Pr}(t_\nu \geq T_o) \leq p \mid t_{\nu, \delta} \right ] 
  & = \mbox{I\!Pr}\left [ \Psi_\nu(T_o) \geq 1 - p \mid t_{\nu,
  \delta} \right ]  \\ & = 1 - \mbox{I\!Pr}\left [T_o \leq \Psi^{-1}_\nu(1-p) \mid t_{\nu, 
  \delta} \right ]  = 1- \Psi_{\nu, \delta}\left[\Psi^{-1}_\nu(1-p)  \right ],
\end{split}
\label{pvalH1}
\end{equation}
using the fact that under the alternative, the test statistic follows
a non-central $t$-distribution with non-centrality parameter $\delta$
and $\nu$ degrees of freedom, and cumulative distribution and
probability density functions $\Psi_{\nu, \delta}(\cdot)$ and
$\psi_{\nu,\delta}(\cdot)$. Letting $\lambda_i$ be the probability
that the $i$th voxel is truly active, and $\delta_i$ as the
voxel-specific non-centrality parameter,
\begin{equation}
\mbox{I\!Pr}(P_i\leq p; \lambda_i, \delta_i ) = (1-\lambda_i)p  + \lambda_i
\left\{ 1-   \Psi_{\nu, \delta_i}\left[\Psi^{-1}_\nu(1-p) \right]\right
\},
\label{cdfPvalue}
\end{equation}
from where it follows upon taking derivatives that the density of
$P_i$ is
\begin{equation}
f_{P_i}(p; \lambda_i, \delta_i) = (1-\lambda_i) + \lambda_i \frac{\psi_{\nu,
  \delta_i}\left[\Psi^{-1}_\nu(1-p)
  \right]}{\psi_{\nu}\left[\Psi^{-1}_\nu(1-p) \right]},\qquad 0 < p < 1.
\label{pdfPvalue}
\end{equation}
The density of the $p$-value at a voxel is thus a mixture of the
standard uniform density and another density involving a parameter 
$\delta_i$. This density, illustrated for $\nu = 122$ and different values
of $\lambda$ and $\delta$ in Figure~\ref{1mixdensity}, is used in our assessment methodology. 

\section{Methods}
\label{methods}
\subsection{Imaging}
\label{imaging}
All MR images were acquired on a GE 1.5 Tesla Signa system equipped 
with echo-planar gradients and using v5.8 software. Structural
$T_1$-weighted images were obtained using a standard spin-echo
sequence with TE/TR of 10/500 ms, and slice-positioning
following the recommendations of Noll et al.~(1997) to minimize
intersession differences. For the fMRI sessions, twenty-four
6~mm-thick slices parallel to the AC-PC line and with no gap between 
them were acquired using a single-shot spiral sequence (with TE/TR of
35/4000 ms)  under a paradigm which involved eight cycles of a simple
finger-thumb opposition task performed for 32s, followed by an equal
period of rest, over 128 time-points. The paradigm was repeated for
twelve separate sessions over a two-month period on a single 
volunteer after obtaining informed consent.  All the paradigms were on
the dominant right hand of the subject. 
Reconstructions were performed on an SGI Origin 200 workstation after
transferring the data from the scanner. Automated image
registration~(AIR) was used to correct for motion-related artifacts
in each replication, after which time series were generated at each
voxel~(Wood et al., 1998) and normalized to remove any linear drift in the
data. Cross-session image registration among the twelve sessions was
additionally performed to minimize any residual misregistration using
the intersession registration algorithms of
AFNI~(Cox and Hyde, 1997). The default first image volume was taken to
be the target against which the images were registered. 
Functional maps were created after computing
voxel-wise $t$-statistics (and corresponding $p$-values) using a
general linear model, discarding the first three image volumes (to
account for $T_1$ saturation effects) and assuming first-order autoregressive
errors, using sinusoidal waveforms  with lags of 8s.  The choice of
waveform represented 
the BOLD response, while the lag duration corresponded to when the
actual BOLD response was seen from the theoretical start of
the stimulus. Finally, 
a composite image cube of $p$-values of activation was created using the
same procedure as above on the combined voxel-wise data from the twelve
replications. (Activation maps as well as true activation and
inactivation certainty measures were computed for this composite image
cube and 
are reported in the Results section.) Since the goal of this
experiment is to detect regions of activation that are positively
associated with the right-hand index finger-thumb opposition motor
task performed by the subject, the alternative is one-sided and hence
the one-sided $t$-tests were used.
The dataset was then transferred to a Dell Precision 650 
workstation, having two 3.06GHz Intel(r) Xeon(tm) processors running
the Fedora 10 2.6.27.9-159 Linux
kernel, where all the algorithms and statistical analyses reported in
this paper were performed using a combination of commands in the ``C''
programming language and the statistical software package {\sf
  R}~(2008)
publicly available for download from {\em  www.R-project.org}.

\subsection{Statistical Methodology}
\label{statistics}
Most statistical analyses of fMRI data involve fitting a (typically
but not necessarily, general linear) model relating the observed time
series at each voxel to the hemodynamic response function~(HRF). A
test statistic is then constructed  and its corresponding $p$-value
is obtained and used 
in identifying activation. In the development here, we use the fact
that the $t$-test is commonly used in analyses; similar
methodology can be developed for the Kolmogorov-Smirnov and other
tests. Further, we use a one-sided $t$-test to illustrate and evaluate
the methodology because our application uses a one-sided 
alternative: we can readily develop similar methodology for two-sided
$t$-tests. 

Let $M$ be the number of replications of the experiment. 
Let $\boldsymbol{ P }_i = \{p_{i,1}, p_{i,2},\ldots, p_{i, M}\}$,
where $p_{i, j}$ is the observed $p$-value of the one-sided
$t$-statistic at the $i$th voxel and the $j$th replication. The
likelihood function for the $i$th voxel is then given by 
\begin{equation}
\prod_{j=1}^M \left\{ (1-\lambda_i) + \lambda_i \frac{\psi_{\nu_j,
  \delta_i}\left[\Psi^{-1}_{\nu_j}(1-p_{i,j})
  \right]}{\psi_{\nu_j}\left[\Psi^{-1}_{\nu_j}(1-p_{i,j}) \right]} \right\}.
\label{llhd.onevoxel}
\end{equation}
The above model assumes that the fixed effect magnitude~(captured
in~$\delta_i$) for each voxel does not vary over the
replications. This assumption is 
similar to that made by the binomial models of Maitra~et
al.~(2002) or Genovese~et al.(1997). However, unlike
the former, the model~(\ref{llhd.onevoxel}) incorporates voxel-specific
probabilities~($\lambda_i$) of true
activation as well as non-centrality parameters~($\delta_i$).
The
degrees of freedom of the test statistic here depend on the
replication, but can also be made voxel-specific, if needed. Under
spatial independence, the likelihood for the entire 
set of voxels in all slices of the corrected image is the
product of~(\ref{llhd.onevoxel}) over all voxels.  

For each voxel, there are two parameters ($\lambda_i$ and $\delta_i$)
to be estimated. Thus, if $N$ is the number of three-dimensional 
image voxels under consideration, there are $2N$ parameters
that are to be estimated from the $MN$ observed $p$-values using the
likelihood model. In our estimation process, we assume that the
observed $p$-values at each voxel are independent with any
spatial relationships fully captured in the voxel-specific $\lambda$s and
$\delta$s. 
Since these are voxel-specific and the observed $p$-values are
independent, maximization can be separately done for each
voxel. This has the benefit of speeding up computation, but has the
limitation mentioned above, namely that the fixed effect magnitude
does not vary over the replications. An alternative approach is to
have replication-specific $\delta$s, rather than voxel-specific fixed
effect magnitudes. Direct maximization would, however, be computationally
impractical then: a possible recourse could be to the
expectation-maximization algorithm of Dempster et al.~(1977). We have
not pursued this course in this paper. 

We use Nelder-Mead's downhill simplex method~(Nelder and Mead, 1965)  to
find the ML parameter estimates~(MLEs). Note that
the likelihood model assumes that $\lambda_i$ is in the interval (0,
1) and that $\delta_i$ is positive so that the parameters are, in
theory, identifiable from the likelihood function. That is, any two
different values of $(\lambda_i, \delta_i)$ in the parameter space 
give rise to different values of the likelihood
function. Numerically, however, small values of $\delta_i$ make
$\lambda_i$ unidentifiable, since the second component density
function is then very close to unity, and any value of $\lambda_i$
yields essentially the same likelihood value. Figure~\ref{1mixdensity}
displays the density function for $\delta=1,2,3$ as the
mixing proportion $\lambda$ increases from 0.05 to 0.95. Note that
when $\lambda=0$ (not pictured), the density is standard uniform, thus
is a horizontal line taking the value 1 for all $p-values$. This is
quite distinct from any of the plotted functions in any of the plots in
Figure~\ref{1mixdensity}. Therefore, we conclude that for $\delta_i >
1$, the two components 
in the mixture of ~(\ref{pdfPvalue}) seem to be well-separated  and
identifiability in estimation does not appear to be a major
issue. Consequently, we restrict $\delta_i > 1$ in our
computations.   

Once these parameter estimates are available, true activation and
inactivation certainty measures of voxels identified as activated and
inactivated can be computed. For let $\tau_i$ be the
threshold at which the $i$th 
voxel is declared to be activated if it has a lower $p$-value. The
thresholds are not necessarily voxel specific and can be assumed to be
obtained by any method, such as those obtained by controlling the False
Discovery Rate~(Genovese et al., 2002) or the related methods surveyed in 
Nichols and Hayasaka~(2003). However, we can also use
methodology~(see below) that maximizes the {\em ML reliability
  efficient frontier}~(Maitra et al., 2002; Genovese et al., 1997)
voxel-wise, which 
also follows from the methodology developed above.
From the threshold values $\tau_i$s, we can compute certainty
measures. To see this, note that the $i$th voxel 
would be identified as activated if its observed $p$-value $P_i$ is
less than or equal to  $\tau_i$. Then the true activation certainty $(\rho^+)$
of this voxel is the posterior probability of a voxel being
truly active given that it has been identified as activated. Using
Bayes' Theorem, 
\begin{equation}
\begin{split}
\rho_i^+ & = \mbox{I\!Pr} (i\mbox{th voxel} \mbox{ is truly active} \mid i\mbox{th voxel is
  identified as activated}) \\
&  = \frac{\mbox{I\!Pr} (i\mbox{th voxel is truly
  active and } p_i < \tau_i)}{\mbox{I\!Pr}(p_i < \tau_i)}\\
& =  \frac{\mbox{I\!Pr} (i\mbox{th voxel is truly
  active})\mbox{I\!Pr}( p_i < \tau_i\mid i\mbox{th voxel is truly
  active})}{\mbox{I\!Pr}(p_i < \tau_i)} = \frac{\lambda_i \left\{ 1- \Psi_{\nu_i,
  \delta_i}\left[\Psi^{-1}_{\nu_i}(1-\tau_i)  \right
  ] \right\}}{(1-\lambda_i)\tau_i  +
\lambda_i  \left\{ 1-   \Psi_{\nu_i,
  \delta_i}\left[\Psi^{-1}_{\nu_i}(1-\tau_i) \right]\right \}} \\
\end{split}
\label{reliability}
\end{equation}
where the numerator follows from~(\ref{pvalH1}) and the denominator
from~(\ref{cdfPvalue}), directly.

In a similar spirit, the {\em true inactivation certainty}~$(\rho^-)$
of a voxel identified as inactive is defined to be 
  the (posterior) probability that it is truly inactive when it has been
  correctly identified as so. Corresponding to the above, this can
  be obtained as 
\begin{equation}
\begin{split}
\rho^-_i & = \mbox{I\!Pr} (i\mbox{th voxel} \mbox{ is truly inactive} \mid i\mbox{th voxel is
  identified as inactivated}) \\
&  = \frac{\mbox{I\!Pr} (i\mbox{th voxel is truly inactive and } p_i\geq\tau_i)}{\mbox{I\!Pr}(p_i\geq\tau_i)}\\
& =  \frac{\mbox{I\!Pr} (i\mbox{th voxel is truly
    inactive})\mbox{I\!Pr}( p_i\geq\tau_i\mid i\mbox{th voxel is truly
    inactive})}{\mbox{I\!Pr}(p_i\geq\tau_i)} \\
& = \frac{(1-\lambda_i)(1-\tau_i)} 
{(1-\lambda_i) (1-\tau_i)  +
\lambda_i  \left\{\Psi_{\nu_i,
  \delta_i}\left[\Psi^{-1}_{\nu_i}(1-\tau_i) \right]\right \}} 
\end{split}
\label{anti-reliability}
\end{equation}
where the numerator and the denominator  follow from
the complement from unity of~(\ref{pvalH0}) and~(\ref{cdfPvalue}),
respectively. 

As indicated above, and although not a focus of this paper, the obtained
parameter estimates can also be used 
with the model to obtain threshold values. To see this, let $\tau_i$
be the given threshold at the $i$th voxel. Then the probability of
making a correct decision by thresholding the $i$th voxel at threshold
$\tau_i$ is equal to 
\begin{equation*}
\begin{split}
\mbox{I\!Pr}(\mbox{correct decision} \!\mid\! \mbox{truly
inactive voxel})\mbox{I\!Pr}(\mbox{truly inactive voxel}) +
\mbox{I\!Pr}(\mbox{correct decision}\!\mid\! \mbox{truly active
voxel})\mbox{I\!Pr}(\mbox{truly active voxel}) 
\end{split}
\end{equation*}
or equivalently, 
\begin{equation}
(1-\lambda_i) (1-\tau_i) + \lambda_i \left\{ 1-   \Psi_{\nu_i,
  \delta_i}\left[\Psi^{-1}_{\nu_i}(1-\tau_i) \right]\right \}.
\label{MLRfrontier}
\end{equation}
The above is also called the {\em ML reliability efficient
  frontier}~(Maitra et al., 2002; Genovese et al., 1997),
and given estimates for $\lambda_i$ and $\delta_i$, we can maximize
the above probability of making a correct decision with respect to
$\tau_i$ to get an optimal threshold. It may be noted that the optimal
thresholds here are also voxel-specific, since the parameters
$\lambda_i$ and $\delta_i$ are so. Hence, different
voxels can be declared as activated at different threshold values,
 providing a data-driven approach to detecting activation after
accounting for spatial context and other inhomogeneities that arise
from the experiment. 

\section{Results}
\label{results}
\subsection{Variability in Activation}
\label{variability}
\begin{figure*}[t]
\includegraphics[height=4.75in, width=7.4in]{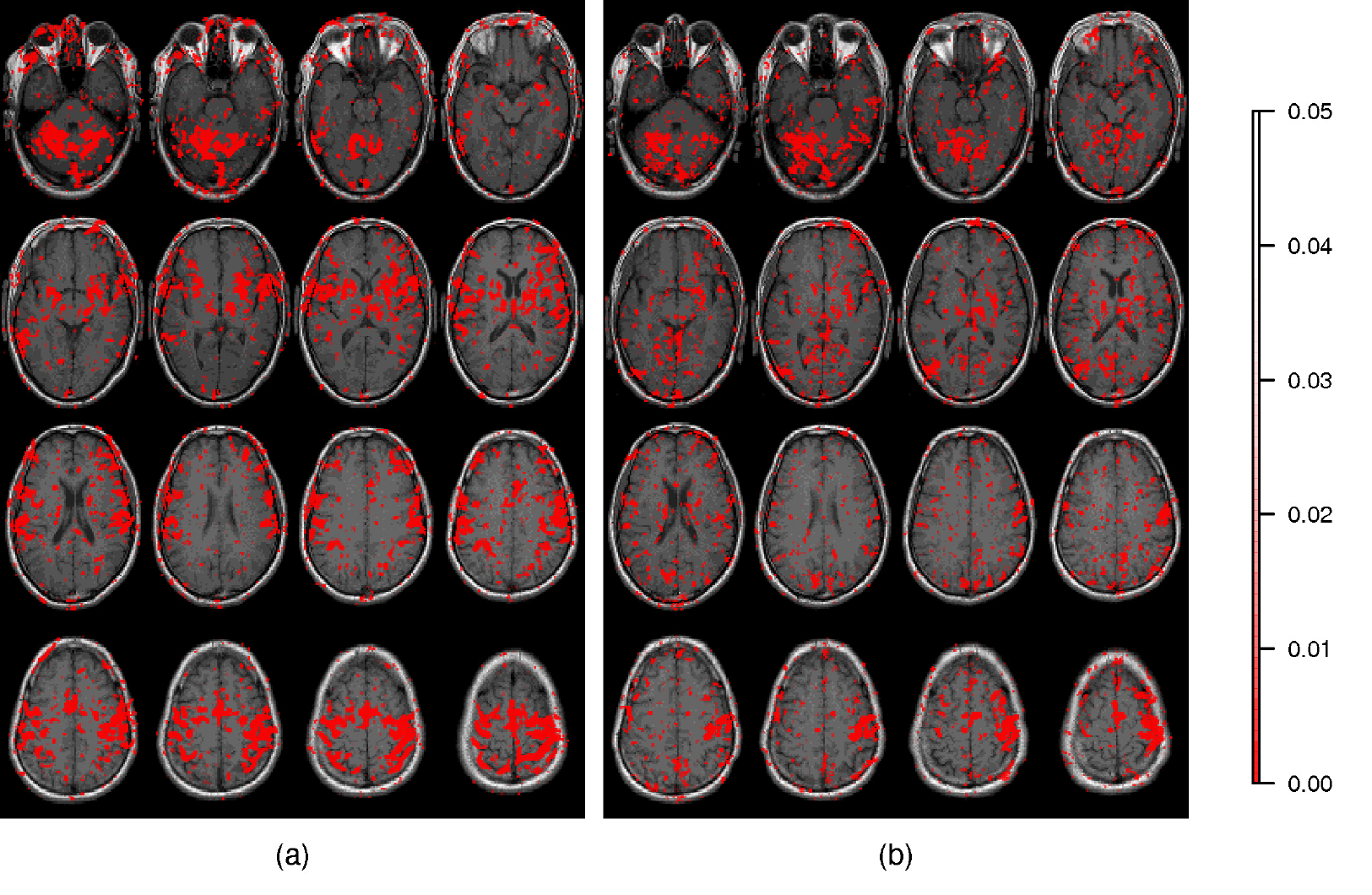} 
\caption{Radiologic view maps of observed $p$-values of activation of
  the $t$-test of motor function for slices 7 through 22 (row-wise) from
  the (a) first and (b) twelfth experiments on a single normal
  subject, overlaid on structural $T_1$-weighted images and using a right
  hand finger-thumb opposition experiment. The opacity of the red
  overlays are inversely proportional to the $p$-value of the
  corresponding $t$-statistic. }
\label{arighthandrep}
\end{figure*}
Figure~\ref{arighthandrep} represents the observed $p$-values of
activation for slices 7 through 22 in the first and last replications
of the experiment. All displays reported in this paper are in
radiologic views and overlaid on top of the 
corresponding $T_1$-weighted anatomical images. Note the
large amount of variability in  the observed $p$-values in between the
two replications. In both cases, the region of the left primary
motor cortex appears to be significantly activated in response to the
task of right finger-thumb opposition. 
But, let us consider, for instance, slice 20~(bottom row, second 
image slice) which shows a substantially large area encompassing the
primary left motor cortex with low $p$-values in the first
replication~(Figure~\ref{arighthandrep}a). There are other large areas
also in this slice which have very low $p$-values. In
Figure~\ref{arighthandrep}b however, the area with low $p$-values in
this region is far more 
concentrated and primarily in the region of the primary left motor
cortex. 

\begin{figure*}
\includegraphics[height=7.4in, width=7.4in]{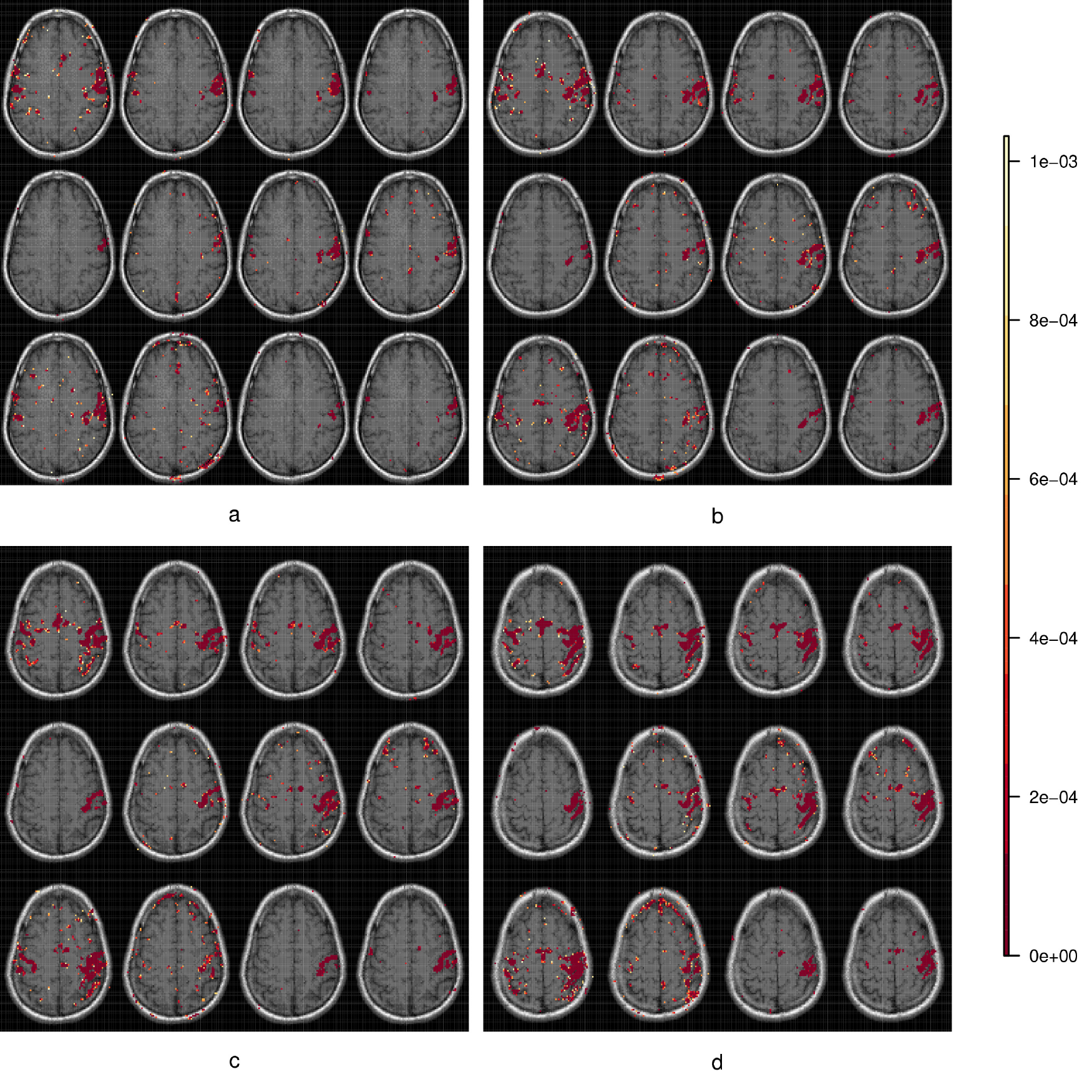} 
\caption{Radiologic view maps for (a) slice 18, (b) slice 19, (c)
  slice 20 and (d) slice 21, of $p$-values for activation regions as
  determined by controlling the False Discovery Rate (FDR) at a
  nominal expected FDR of $q$=0.05. For each slice, we display the
  $p$-values of activation for the thresholded voxels using a $t$-test of
  the motor function for the twelve replications of the right hand
  finger-thumb opposition experiment on the same volunteer. Note the
  differences in location and extent of activation over the twelve
  replications.
}
\label{fdrrel}
\end{figure*}
\begin{figure*}[t]
\includegraphics[height=3.7in, width=7.4in]{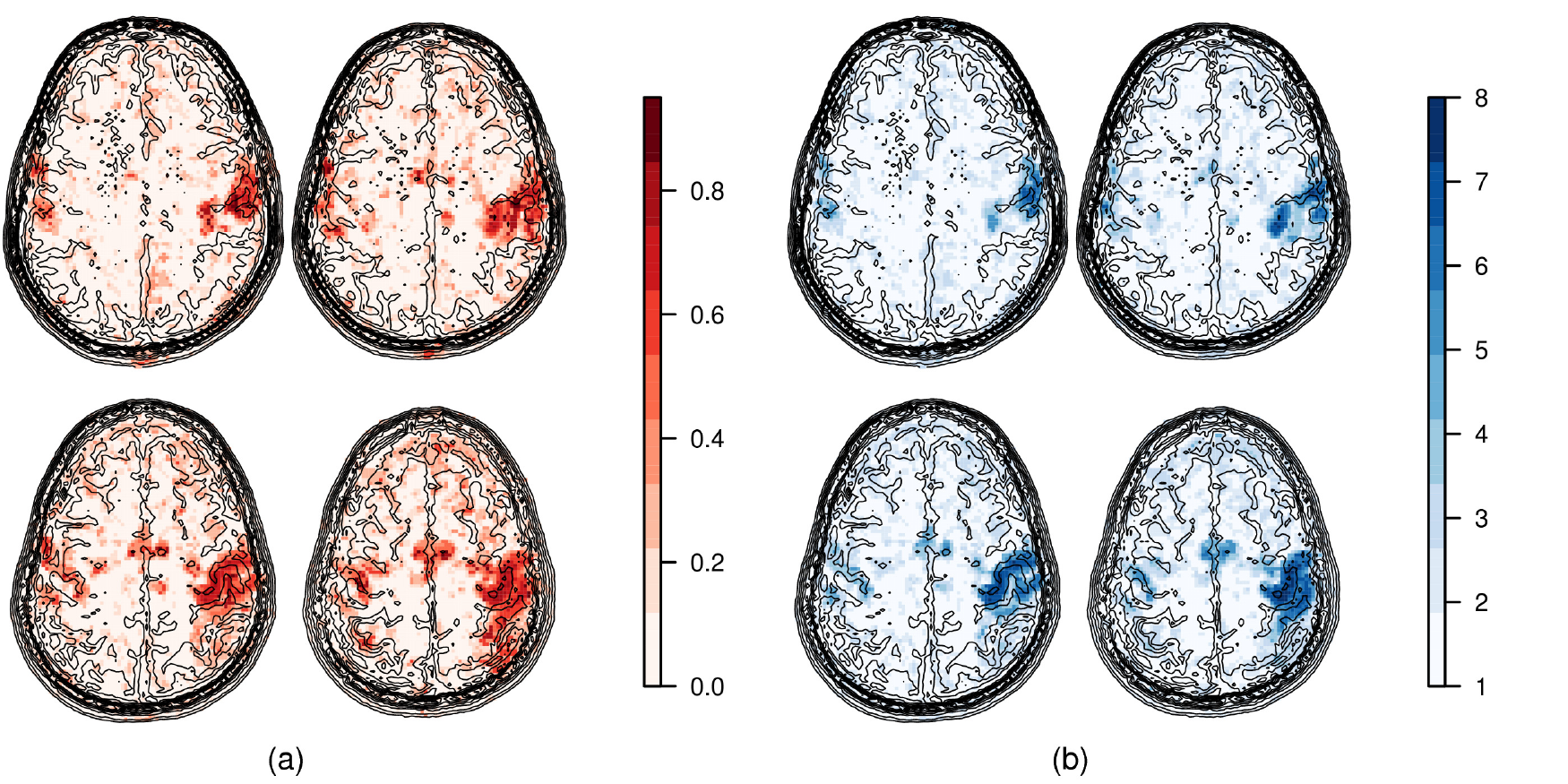} 
\caption{Estimated (a) $\lambda$ and (b) $\delta$ images for slices
  18, 19, 20 and 21 (row-wise) in the right hand finger-thumb
  opposition experiment. A contour plot of anatomic detail of each
  slice is overlaid on the corresponding image.}
\label{lambdancp}
\end{figure*}
These two figures illustrate the across-session variability  in
observed $p$-values for the same paradigm on the same subject. This 
variability can impact the results of experiments and scientific
conclusions. To see this, consider the results of using the 
Benjamini and Hochberg~(1995) approach to determining
activation by controlling the expected false discovery rate~(FDR)
nominally at $q=0.05$ separately, for each of the  twelve
experiments. Figure~\ref{fdrrel} displays 
radiologic views of the $p$-values of voxels determined as activated
in the eighteenth through the twenty-first slices encompassing the
ipsi- and contra-lateral pre-motor cortices~(pre-M1), the primary motor
cortex~(M1),  the pre-supplementary motor cortex (pre-SMA), and the
supplementary motor cortex~(SMA). Clearly, there is wide variability
in the results. Thus, while all experiments identify activation in the
left M1 and in the ipsi-lateral pre-M1 areas, there is wide
variability in identified activation in the contra-lateral pre-M1,
pre-SMA and SMA voxels, with some experiments (most 
notably, the fifth, eleventh and to a lesser extent, twelfth
replications) reporting very localized or no activation, while in
other cases, these areas are identified as activated and indeed, the
identified activated regions are sometimes more diffused. Indeed, the
66 $R_{jm}$s range from 0.081 to 0.494, with a median value of 0.228
and a inter-quartile range of 0.115. Figure~\ref{fdrrel} illustrates
the need for variability assessment 
very nicely. Conclusions based on any of the twelve replications that 
do not account for the variability in the experiment could be very
different and potentially erroneous. Hence, some quantification of
variability in the observed activation is needed. We demonstrate
use of our methodology towards this goal in the next section.

\subsection{Illustration of Methodology}\label{illustration}

Our Nelder-Mead minimization routines for the converged ML estimates
of $\lambda_i$s and $\delta_i$s at each voxel took around 5
milliseconds. Thus calculations on the estimated parameters for the
entire set of images took a little  more than half an
hour. Figure~\ref{lambdancp} displays the estimated $\lambda$s and
$\delta$s for slices 18, 19, 20 and 21. Note that we 
process and estimate parameters for all slices, but henceforth only
display these four slices for clarity of presentation. 
The $\lambda$-values are voxel-wise estimates of 
probability of true activation and it is encouraging to note from
Figure~\ref{lambdancp}a that they are high in known regions of
activation such as the left M1, the ipsi- and contra-lateral pre-M1
areas, and also moderately so in the pre-SMA and SMA areas. 
Similar trends are also reported for the estimated
$\delta_i$s~(see Figure \ref{lambdancp}b).
\begin{figure*}[t]
  \begin{minipage}[t]{0.47\linewidth}
  \includegraphics[width=\textwidth]{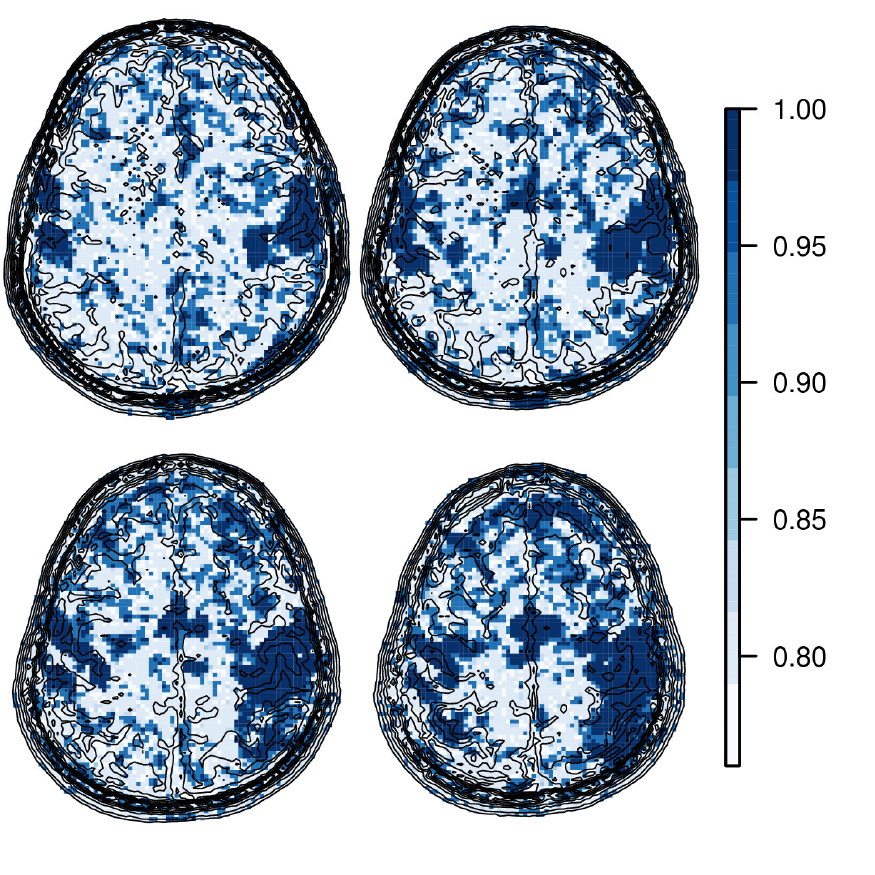}
\caption{Area under the voxel-wise estimated receiver operating
  characteristic (ROC) curves for slices 18, 19, 20 and 21 for the
  right hand finger-thumb opposition experiment.}
\label{righthand20thsliceROC}
  \end{minipage}\hfill
  \begin{minipage}[t]{0.47\linewidth}
\includegraphics[width=\textwidth]{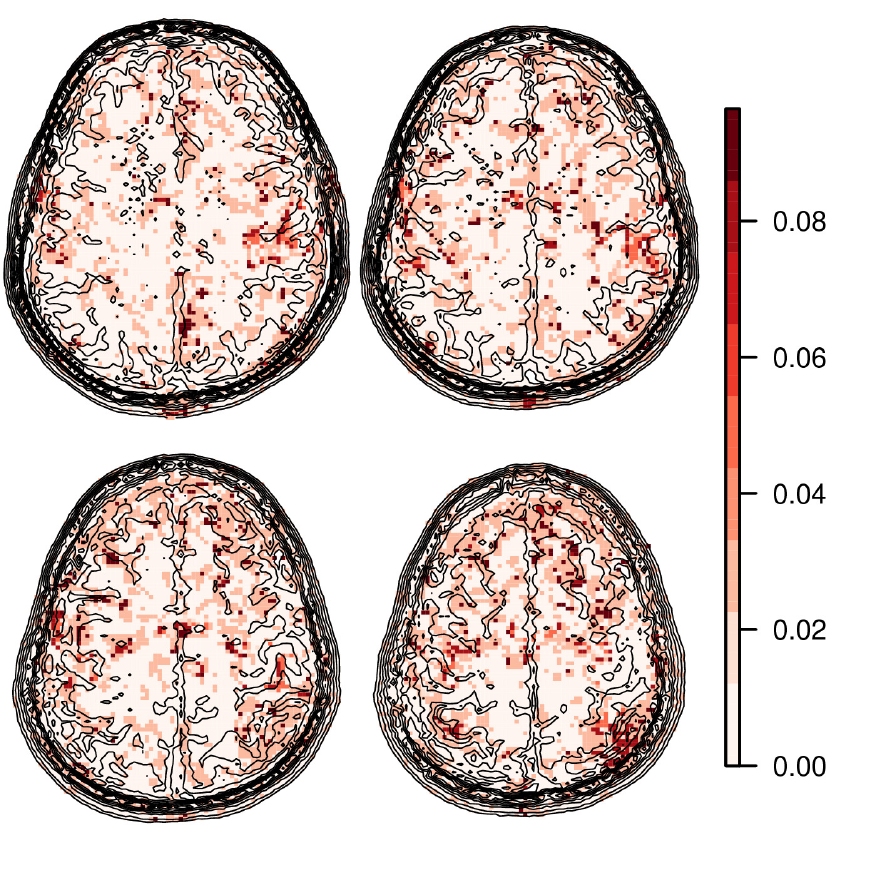}
\caption{Optimal threshold values maximizing the reliability efficient
  frontier for slices 18, 19, 20 and 21 (row-wise) for the right-hand
  finger-thumb opposition experiment.}
\label{righthandcutoffs}
 \end{minipage}
\end{figure*}
Unlike in the setup of Genovese et al.~(1997) or
Maitra et al.~(2002), every voxel has an individual ROC
curve. Alternatively, the probabilities of true positives and false
negatives ($\pi_A$ and $\pi_I$) take different values for the same
thresholds at different voxels. For instance, at the threshold
$\tau_j$, $\pi_I$ is also $\tau_j$ regardless of voxel while
from~(\ref{pvalH1}), we get 
${\pi_A}_i = 1- \Psi_{\nu, 
  \delta_i}\left[\Psi^{-1}_\nu(1-\tau_j)\right ] $ for the $i$th
voxel. Figure~\ref{righthand20thsliceROC} summarizes the ROC
voxel-wise in terms of the area under the curve~(AUC) for the four
slices. The AUC is an average of the sensitivity over all possible
specificities~(Swets, 1979; Hanley and McNeil, 1982; Metz, 1986), with
high values 
indicating good discrimination between truly 
activated and inactivated voxels. Thus, it is encouraging to 
note that the AUCs are very high in areas such as the left M1
that are known to have a high chance of true activation~($\lambda_i$),
providing confidence in the results derived using our modeling approach.   

\begin{figure*}[h]
\includegraphics[height = 4.1in, width = 7.2in]{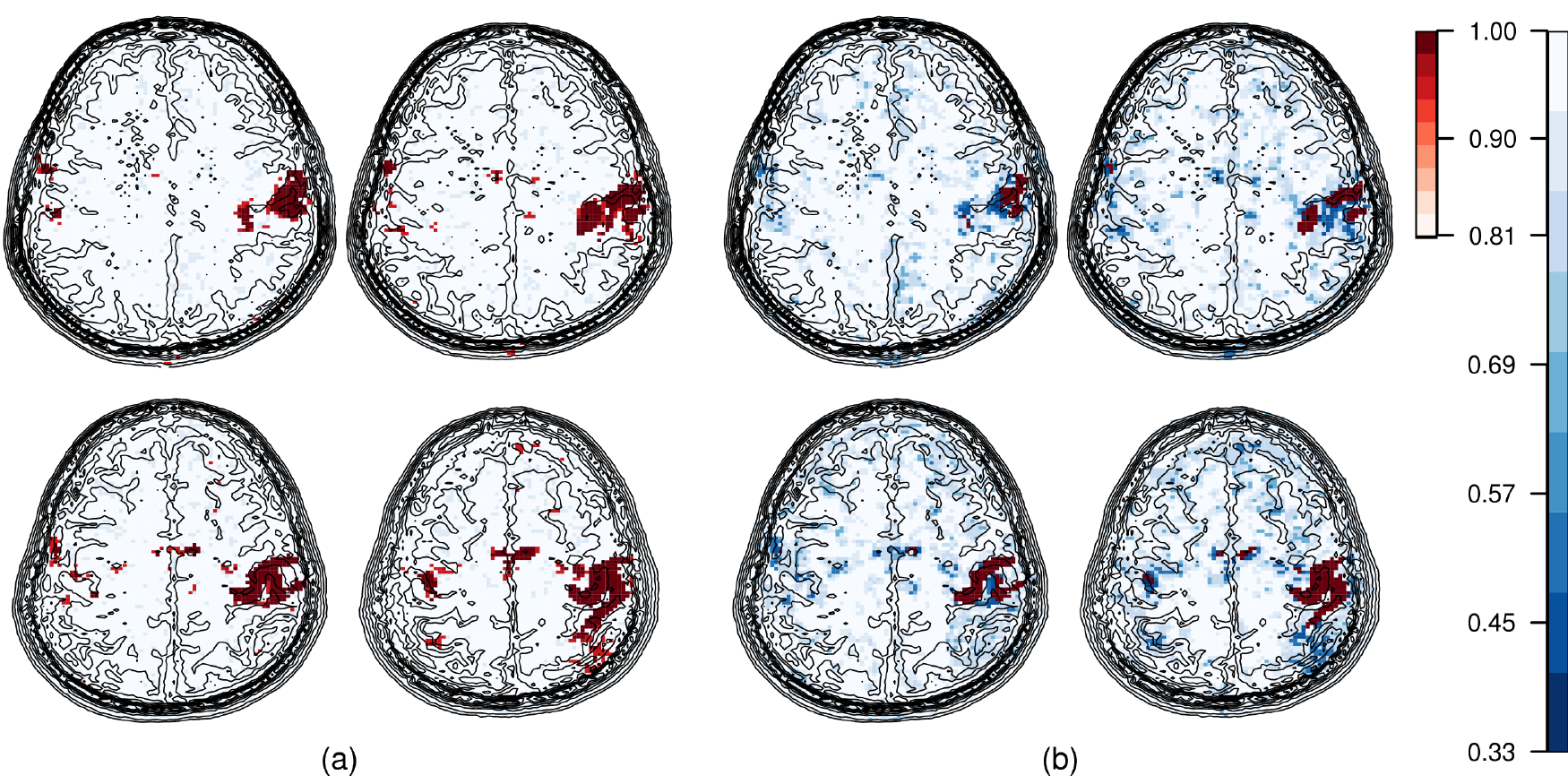}
\caption{Activation maps on the composite image from twelve
  replications obtained using the
cutoffs derived from (a) maximizing the reliability efficient frontier
and (b) the expected false discovery rate, along with their
corresponding true activation (in red) and true inactivation certainties
(in blue) for the right-hand finger-thumb opposition experiment. }
\label{righthandreliabilities}
\end{figure*}
Figure~\ref{righthandcutoffs} displays the derived voxel-wise cutoffs
optimizing the ML reliability efficient of~(\ref{MLRfrontier}) for the
four slices. These are the thresholds at which our confidence in the
reliability of detected activation is the greatest. The figures
indicate that our confidence of a correct activation is highest for
truly active voxels (such as left M1) at higher
thresholds whereas very low thresholds are required for greater
accuracy in other regions. 

Comprehensive activation maps were obtained by thresholding the
$p$-values of the composite image with the voxel-wise thresholds of
Figure~\ref{righthandcutoffs}. These maps along with their $\rho^+$
and $\rho^-$ values are displayed in
Figures~\ref{righthandreliabilities}. Note that the $\rho^-$-values 
for this figure are quite high, corresponding to a lowest value of
0.62. We also calculated the true certainty
measures~(Figure~\ref{righthandreliabilities}b) for the areas of  
activation and inactivation identified on the composite image by the
FDR thresholding methods~($q=0.05$) of Benjamini and
Hochberg~(1995)
as adapted to fMR data by  Genovese~et al~(2002). (Results
obtained using the more liberal approach of Storey and
Tibshirani~(2003) were essentially the same as in
Figure~\ref{righthandreliabilities}b  and are not displayed.)
These two figures very nicely illustrate the value of certainty
assessment and also the performance of our method in the context of
threshold-identification and certainty calculations. To see
this, note that the areas identified as activated by FDR, such as the
left ipsi-lateral pre-M1 and left M1 areas are a  
subset of those that are identified as 
activated using the thresholding of our method. This is encouraging
because FDR methods are known to be overly conservative when a large
number of null hypotheses are known to be true, as is the case in fMRI
experiments where most voxels are known to exhibit no activation. At
the same  
time, voxels in the SMA, pre-SMA and contra-lateral pre-M1
regions that are identified as activated using our thresholding 
but inactivated using FDR have high $\rho^+$ and $\rho^-$ values
respectively. This is very encouraging because this means  
that the more conservative FDR method has missed areas of activation
(such as in the pre-SMA, SMA, and contra-lateral pre-M1 regions) that  
our thresholding picked up with correspondingly high $\rho^+$-values but
these have low $\rho^-$-values under the FDR thresholding. Thus
even though these areas were not identified as activated by FDR, they
have a good chance of being truly active, illustrating the value of
assessing certainty in the results using our estimation
method. Further, even   
though different thresholdings are used in the two approaches of 
Figures~\ref{righthandreliabilities}a and b, yielding different
values for $\rho^+$ and $\rho^-$, the results are
consistent. 
Thus, for any fMR experiment with test-retest data, we can not only 
obtain an activation map, but also a detailed map of the true activation
and inactivation certainties of voxels that are identified as activated or 
inactivated, providing a tool for the investigator to quantify
results. 

\subsection{Assessment of Methodology}
\label{assessment}
The methodology was evaluated through a series of numerical
experiments performed by generating $M$ replicated three-dimensional
images of simulated $p$-values using the density in~(\ref{pdfPvalue}) and with
the $\lambda_i$s and $\delta_i$s estimated from the above dataset as
\begin{wraptable}{r}{0.5\linewidth}
\vspace{-0.15in}
\caption{RMSEs of estimated $\hat\lambda$s (left column) and
  $\hat\delta$s (middle column) for the simulation experiments for
  different replication sizes using our method. Squared
  Hellinger distances (SHD) averaged over all voxels  between the estimated
  densities and the ``ground   truth'' densities are provided in the
  third column.}
\label{table1}
\begin{tabular}{|c|c|c|c|}\hline
{\bf \# replications} & {\bf RMSE($\hat\lambda$)} & {\bf
  RMSE($\hat\delta$)} &{\bf Average SHD}\\ \hline
2 & 0.239 & 2.220 & 0.092 \\
3 & 0.222 & 2.394 & 0.068 \\
4 & 0.237 & 2.597 & 0.061 \\
5 & 0.235 & 2.690 & 0.062 \\
6 & 0.223 & 2.554 & 0.052 \\
7 & 0.224 & 2.633 & 0.055 \\
8 & 0.234 & 2.731 & 0.042 \\
9 & 0.235 & 2.783 & 0.042 \\
10 & 0.242 & 2.854 & 0.039 \\
11 & 0.244 & 2.887 & 0.036 \\
12 & 0.224 & 2.677 & 0.035 \\ \hline
\end{tabular}
\vspace{-0.35in}
\end{wraptable}
the ``ground truth''. Our methodology was 
then used to estimate the parameters of the model. Estimation performance 
was assessed in terms of the Root Mean Squared
Error~(RMSE) of the estimated $\hat\lambda_i$s and $\hat\delta_i$s
obtained using our methodology on the simulated
data. Formally, RMSE~($\hat\lambda$) $=\sqrt{\sum_{i=1}^N (\hat\lambda_i
  -\lambda_i)^2/N}$ while  RMSE~($\hat\delta$) $=\sqrt{\sum_{i=1}^N
  (\hat\delta_i 
  -\delta_i)^2/N}$. Further, since the estimated parameters impact
performance of our methodology together and through the density, we
calculated the squared Hellinger distance between the two densities: 
$\int_0^1  [\sqrt{f_{P_i}(p; \hat\lambda_i, \hat\delta_i)} -
  \sqrt{f_{P_i}(p; \lambda_i, 
  \delta_i)}]^2 dp$ where $f_{P_i}(\cdot)$ is as
in~(\ref{pdfPvalue}). We repeated the process for $M = 2,3,\ldots,
12$ to assess how performance changes with different numbers of
replications. Performance measures on the  RMSEs and the averaged
Hellinger distance over all 
the voxels are in Table~\ref{table1}. Note that while the RMSEs are
modest for $\hat\lambda$s, they are somewhat higher for the
$\hat\delta$s. However, the squared Hellinger distance averaged over all
voxels is quite low, pointing to good performance of the
methodology. We note, however, that performance with only two
replications is not very good, and perhaps at least three replications
are needed. Interestingly, it is a bit unclear whether the RMSEs  go
down consistently with increasing number of 
replications. This potentially points to some ill-posedness in the estimation
process -- a view that is further strengthened by noting that the
RMSEs and squared Hellinger distance measures are heavily inflated by
a few scattered voxels. Hence, incorporating some amount of
regularization through a penalty function on the $\lambda_i$s and the
$\delta_i$s as in Maitra~et al.~(2002) may be appropriate. 

\begin{figure*}
\includegraphics[height = 4.1in, width = 7.2in]{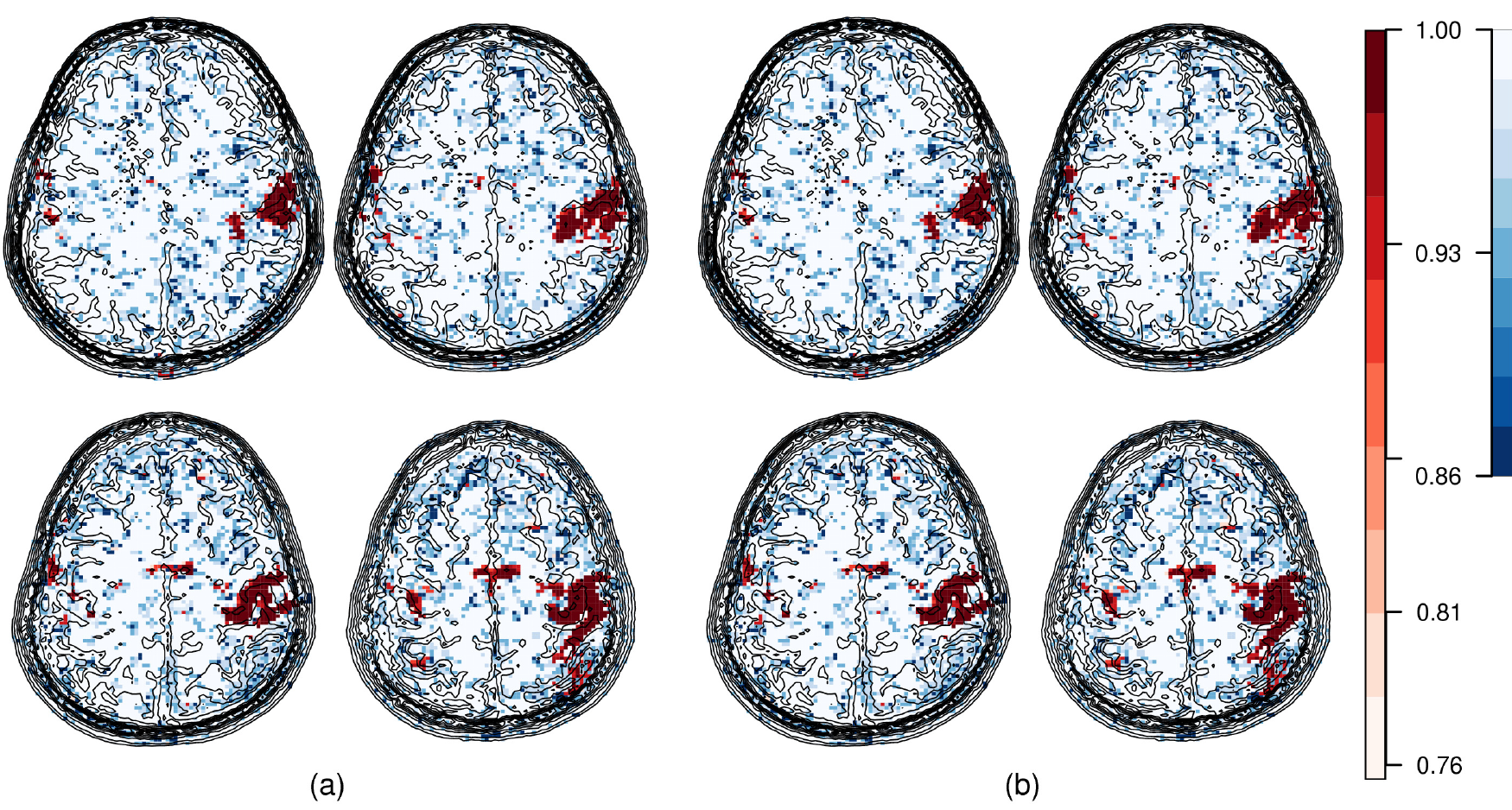}
\caption{Activation maps on slices 18, 19, 20 and 21 (row-wise) of the
  composite image from 
twelve replications with corresponding true activation (red) and
inactivation certainties (blue) obtained from parameters and thresholds
estimated using (a) a randomly chosen sample of six replications from
the data and (b) the other six replications, complementary to the set
in (a). Displays are as in Figure~\ref{righthandreliabilities}.}
\label{righthandrobustness}
\end{figure*}
The robustness of the methodology in detecting activation and the
certainty measures was also evaluated. The twelve replications in
the dataset were randomly subdivided into two groups of six each, and
our methodology was applied to each subset to obtain estimated
parameters, as well as the
thresholds maximizing the ML reliability efficient frontier. These
were used separately to compute the $\rho^+$ and $\rho^-$ measures of the
composite image map of 
$p$-values. Figure~\ref{righthandrobustness} shows these maps using
the two random subsets of six replications each. It is encouraging to
note that the activation maps as well as the true certainty  measures are
essentially the same for both Figures~\ref{righthandrobustness}a and
b, pointing to  robustness of the suggested methodology in detecting
activation.  

\section{Discussion}
\label{discussion}
Genovese et al.~(1997) and Maitra et al.~(2002)
provided novel approaches to estimating the test-retest certainty of
a voxel using ML and its penalized version to enforce
spatial dependence between the estimated parameters. In both cases,
the approach needs some processing by thresholding the acquired fMRI
data before the models can be applied. The number of threshold levels
and the thresholding values are subjective and depend entirely on
the investigator. This paper removes the need for this step by
modeling the $p$-values of activation directly as a mixture of two
distributions --  one under the null hypothesis of no activation, and the
other under the alternative hypothesis of true activation. Most fMRI
data are processed using $t$-statistics obtained after fitting a
general linear model, and we illustrate our methodology under this
setup. We use this model and ML methodology to estimate
the voxel-wise probability of true activation, and also other model
parameters such as the 
non-centrality parameter which is allowed to be voxel-specific in
order to account for systematic variations owing to local
inhomogeneities  in the magnetic field. These estimated parameter 
values can be easily used to obtain optimal thresholding values in
order to  determine if  a voxel is activated or
inactivated. True activation and inactivation certainty measures of the
activated and inactivated voxels can then be calculated and used by
the investigator to obtain a quantitative assessment of the extent of
activation. Voxel-specific ROC curves were also obtained for each
voxel. Finally, the method was evaluated for its estimation
performance and also for robustness in detecting and
quantifying certainty of activation.

Two reviewers have very kindly asked about the practical utility of
the derived methodology. This paper demonstrates certainty calculations on
replicated single-subject experimental data. The end result is an
individual activation map along with corresponding certainty measures
of activation and inactivation. This provides understanding and
quantitation of the activation of single-subject brains, which is
important for clinical purposes. The methodology can also be used in
the context of replicated and non-replicated data on the same
experimental task or condition performed by multiple subjects. For
each subject, one would draw an activation map and calculate the
individual certainty measures of activation and inactivation. Once
again, individual certainty measures for each subject could
potentially be useful for clinical diagnosis: for instance one may be
interested in finding out reasons for an individual's low certainty
measures of activation/inactivation in understanding how his brain
compares with the rest. These measures can provide the researcher and
the neurologist with a starting point for clinical  investigation and
diagnosis.  

The certainty measures estimated in this paper were dependent
entirely on the statistical analysis chosen to prepare the activation
maps. Thus, it is imperative that fMR data are adequately cleaned and
post-processed before analysis. For instance, one may have 
draining veins in an area as determined by an MR angiographic scan. In
this case, an appropriate approach would be to mark the voxels in this
region as inactive and use this additional information in the modeling and
estimation. Other more sophisticated analysis such as in Saad et
al.~(2001) may also be considered. Further, data may also be
digitally filtered~(Genovese et al., 1997)  prior to analysis in order to
account for 
physiological factors such as cardiac and respiratory motion which
greatly degrade the quality of activated maps. 

There are a number of other remaining issues  that merit further
attention. In the derivations and analysis in this paper, we ignored
any spatial structure among the parameter values in the estimation
process. Approaches such as in Maitra~et al.~(2002) can be
easily incorporated in the model and are a natural extension. This
would also allow for incorporating smoothness that is introduced, 
as kindly suggested by a reviewer, in the registration step of
pre-processing. This would also help in reducing the number of
replications needed, and also in providing statistical consistency in
the estimates, as mentioned in Section~\ref{assessment}.
Further,
the methodology suggested in this paper was developed using
$t$-tests. One advantage of the thresholding approach of Genovese et
al.~(1997) is that replications analyzed using different
testing strategies could be analyzed together using very little
additional effort. Though our entire development here used the most
commonly used $t$-tests,  our methodology is general
enough to be modified and extended to situations involving other kinds
of analysis 
(such as Kolmogorov-Smirnov tests), or when the replicates are
analyzed using different testing strategies. In this case, the
model underlying these other testing strategies will need to be 
explicitly incorporated in the development. 

A separate issue involves applicability of this methodology to grouped
fMRI data, such as in Gullapalli et al.~(2005). It would
be illustrative to see  how certainty of activation/inactivation with
grouped data using  
our suggested method compares with that done in that paper. One could
also compare with the other traditional measures of reliability for
grouped data, such as the ICC. Further
investigations are also needed in order to test the utility of the
methodology to studies done using other paradigms. Finally, one issue of
great interest to researchers in cognitive sciences is to determine
the certainty of activation maps obtained from a single-session
study. In many cases, the nature of the experiment makes it impossible
to have more than one session (hence replication) to acquire fMRI
data. The test-retest methodology derived in this paper is
inapplicable in such situations, and there is therefore great need for
similar methods for such a scenario. One possibility is to model the
runs, each of which occurs when a task is performed during a single fMRI visit
or replication.  Typically, multiple runs occur within the same
replication. Note that runs necessarily have a dependence structure
between them which will need to be modeled.
Thus, while this paper introduces promising methodology to assessing
certainty in test-retest fMRI activation studies, a number of issues
remain that merit further attention.  
\section{Acknowledgments}
\label{acknowledgments}
I thank Rao P. Gullapalli of the University of Maryland School of
Medicine for providing me with the data used in this study and for
many other helpful discussions related to this paper. This material is based,
  in part, upon work supported by the National Science
  Foundation~(NSF) under   its CAREER Grant No.  DMS-0437555 and 
by the National Institutes of Health~(NIH) under its Grant
No. DC-0006740.  

\end{document}